\documentclass[oneside,twocolumn,preprintnumbers,amsmath,amssymb,nofootinbib,showpacs]{revtex4}

\usepackage{amsmath}
\usepackage{amsfonts}
\usepackage{amssymb, multirow}
\usepackage{graphicx}
\usepackage{array}
\usepackage[colorlinks=true,linktocpage=true,linkcolor=blue,citecolor=red]{hyperref}

\usepackage{cleveref}
\usepackage[utf8]{inputenc}
\crefname{section}{§}{§§}

\begin{document}

\title{Single-step de Sitter vacua\\[2mm]from non-perturbative effects with matter\\[2mm]}

\author{Adolfo Guarino and Gianluca Inverso}

\affiliation{
\vspace{5mm}
Nikhef, Science Park 105, 1098 XG Amsterdam, The Netherlands
}

\begin{abstract}

A scenario of moduli stabilisation based on the interplay between closed and open string sectors is explored in a bottom-up approach.
We study ${\mathcal{N}=1}$ \mbox{effective} supergravities inspired by type~IIB orientifold constructions that include background fluxes and \mbox{non-perturbative} effects.
The former generate the standard flux superpotential for the axiodilaton and complex structure \mbox{moduli}.
The latter can be induced by gaugino condensation in a non-Abelian sector of D7-branes and involve the overall K\"ahler modulus of the compactification as well as matter fields. We analyse the dynamics of this coupled system and show that it is compatible with single-step moduli stabilisation in a metastable de Sitter vacuum. 
A novelty of the scenario is that the F-term potential suffices to generate a positive cosmological constant and to stabilise all moduli, except for a flat direction that can be either lifted by a mass term or eaten up by an anomalous U(1).



\end{abstract}

\pacs{04.50.-h, 04.65.+e, 11.25.Mj, 98.80.-k \\[1mm]NIKHEF 2015-044 \\[1mm] e-mails: aguarino@nikhef.nl , ginverso@nikhef.nl}

\maketitle

\section{Introduction}
\label{Sec:Intro}

The fact that our Universe is currently undergoing a phase of accelerated expansion makes the search for long-lived metastable de Sitter (dS) vacua a key step towards linking strings to cosmological data. In the last $10\,$-$15$ years, a lot of progress has been made in the issue of moduli stabilisation in dS vacua. The problem can be tackled both by a string-theoretic approach, constructing explicit backgrounds that give rise to effective actions with desirable properties, and from a bottom-up point of view, identifying models that capture general features of string constructions and solving them in search for dS solutions. In this letter we will adopt the latter approach.

In the KKLT mechanism \cite{Kachru:2003aw}, the type IIB axiodilaton $S$ and complex structure modulus $U$ are assumed to be stabilised at a high scale due to non-trivial Ramond-Ramond (RR) and Neveu Schwarz-Neveu Schwarz (NSNS) background fluxes threading the internal space \cite{Giddings:2001yu}. The outcome of the $S$ and $U$ stabilisation is a constant contribution $\,W_{0}\,$ to an effective  ${\mathcal{N}=1}$ superpotential $\,W\,$ which also incorporates non-perturbative effects coming from D-brane instantons or gaugino condensation on a stack of D7-branes \cite{Derendinger:1985kk,Dine:1985rz}. The latter introduce a further dependence on the K\"ahler modulus $\,T\,$ which results in a superpotential of the form
\begin{equation}
\label{W0+Wnp}
W(T) = W_{0} + A \, e^{i \, a \, T}  \ ,
\end{equation}
where $\,A\,$ and $\,a\,$ are constants.
With an appropriate tuning of the parameters $W_{0}$, $A$ and $a$, the superpotential in (\ref{W0+Wnp}) stabilises the overall  K\"ahler modulus $\,T\,$ in an anti-de-Sitter (AdS) vacuum, fixing the size of the compactification space.  
Some uplift mechanism is then necessary to obtain a de Sitter vacuum.
In the original KKLT scenario such an uplift was proposed based on anti-D3-branes.
Alternative uplift mechanisms based on D-terms \cite{Burgess:2003ic,Villadoro:2005yq} -- possibly combined with perturbative K\"ahler corrections to stabilise the $\,T\,$ modulus \cite{Balasubramanian:2004uy,vonGersdorff:2005bf,Berg:2005yu,Parameswaran:2006jh,Westphal:2006tn,Rummel:2011cd} -- were put forward soon after.

In the D-term uplifting scenario, an open string sector in the form of a matter field $M$ is required to enter the non-perturbative gaugino condensation superpotential \cite{Burgess:2003ic}.
This scenario was put in a gauge-invariant context by Ach\'ucarro et al. (ACCD) in \cite{Achucarro:2006zf} (see also \cite{Haack:2006cy}).
The D-terms were shown to originate from an anomalous $\textrm{U}(1)_{X}$ factor in the U$(N_c)=\mathrm{U}(1)_X\times\textrm{SU}(N_c)$ gauge theory on a stack of $N_c$ D7-branes, where the anomaly is canceled a la Green-Schwarz \cite{Green:1984sg} by the axion component of the $T$ modulus. The stabilisation of $\,S\,$ and $\,U\,$ in ACCD is still assumed to happen at a high scale. The resulting superpotential now carries a dependence on the K\"ahler modulus $T$ and the matter field $M$ of the form 
\begin{equation}
\label{W0+Wnp_D-term}
W(T,M) = W_{0} + A(M) \, e^{i \, a \, T} \ .
\end{equation}
The F-term scalar potential coming from (\ref{W0+Wnp_D-term}) is then combined with a positive definite \mbox{D-term} contribution that depends on the $\textrm{U}(1)_{X}$ charges (see sec.~\ref{Sec:D-terms}). Upon adjustment of the charges and of the $W_{0}$ parameter, a metastable dS vacuum was achieved in \cite{Achucarro:2006zf}.

More recently, the dynamics of type IIB closed string moduli when background fluxes and non-perturbative effects are considered \textit{simultaneously} has been explored \cite{Blaback:2013qza,Kallosh:2014oja,Marsh:2014nla}. 
We will refer to these theories as $\mathcal N=1$ STU-models, based on superpotentials of the form
\begin{equation}
\label{W_STU}
W(S,T,U) = W_{\textrm{flux}}(S,U) +  A(S,U) \, e^{i \, a \, T} \ .
\end{equation}
The pre-factor $A(S,U)$ of the non-perturbative contribution to the superpotential in general depends on the axiodilaton and complex structure moduli \cite{Witten:1996bn}. It was taken to be of polynomial type in \cite{Blaback:2013qza} relying on duality arguments. Interestingly, metastable de Sitter vacua were found where moduli stabilisation occurs as a single-step process. In other words, there is no necessity to invoke an uplift mechanism as the dS solution is obtained from the pure F-term potential.

The possibility of obtaining pure F-term de Sitter vacua with single-step moduli stabilisation has several advantages, including avoiding certain possible consistency issues with the two-step process \cite{deAlwis:2005tf}.
However, the search for such vacua can be computationally difficult and often several simplifying assumptions are made, for instance on the pattern of supersymmetry breaking, in order to reduce the amount of independent variables that come into play. More dS solutions might therefore hide in those regions of parameter space where these simplifications do not hold. Another fact to take into account is that non-perturbative contributions arising from \mbox{D-brane} instantons or gaugino condensation often include a dependence on matter fields, associated with open string degrees of freedom, which cannot be ignored for minimisation issues. It is therefore important to investigate whether single-step metastable dS vacua occur in scenarios also including such matter contributions.

In this letter we tackle this problem from a bottom-up point of view, considering a minimal setup to model the interactions between STU moduli and matter fields. The resulting models, to which we refer as STU$|$M-models, are defined in terms of a superpotential of the form
\begin{equation}
\label{Superpotential_intro}
W(S,T,U,M) = W_{\textrm{flux}}(S,U) +  A(M) \, e^{i \, a \, T} \ .
\end{equation}
When $A(M)$ is homogeneous in the matter field, there is a natural flat direction in the F-term induced potential.
This can be either lifted by an explicit mass term or eaten up by an anomalous U$(1)_X$.
We will look at the full four (complex) field dynamics without assuming stabilisation of $S$ and $U$ at a higher scale.
Instead, by direct minimisation of the scalar potential, we will present examples of single-step moduli stabilisation in a metastable de Sitter vacuum and discuss their main features.

\section{STU$|$M-models}
\label{Sec:model}

The $\mathcal{N}=1\,$ \mbox{STU$|$M-models} we will look at are specified in terms of a K\"ahler potential of the form \cite{Achucarro:2006zf}
\begin{equation}
\label{Kahler}
\begin{array}{lll}
K &=& - \log [ -i (S-\bar{S}) ] - 3 \log [ -i (U-\bar{U}) ] \\[2mm]
& &  - \, 3 \log [ -i (T-\bar{T}) ] + N_{f} M \bar{M} \ .
\end{array}
\end{equation} 
The perturbative sector of this model matches the $\mathbb{T}^{6}/\mathbb{Z}^{2}_{2}$ orientifold, restricted to the isotropic sector \cite{Derendinger:2004jn}, which we regard as a toy model of flux compactifications.
With this interpretation in mind, the $\,S\,$ and $\,U\,$ chiral fields are the type IIB axiodilaton and complex structure moduli, respectively, whereas $T\,$ is the K\"ahler modulus parameterising the overall size of the internal space.
We take a tree-level K\"ahler potential for the closed string moduli $\Phi$ of the form $K \sim - \log [ -i (\Phi-\bar{\Phi}) ]$.
The field $M$ is not part of the closed string moduli of the $\mathbb{T}^{6}/\mathbb{Z}^{2}_{2}$ orientifold.
We introduce it to model interactions with matter fields, possibly coming from open string sectors, induced by non-perturbative effects.
These fields are taken with a flat K\"ahler potential \cite{Haack:2006cy,Lebedev:2006qq,Achucarro:2006zf}\footnote{Closed string moduli typically modify the canonical structure of the K\"ahler potential for the matter fields (see \textit{e.g.} \cite{Ibanez:1998rf}). However, such a modification depends considerably on the specifics of the string embedding of the effective model, thus going beyond the scope of this letter. It is therefore important to assess the robustness of the de Sitter vacua presented in this work with respect to this and other string theoretic modifications of the scenario. We will come back to this in the discussion section.} and appear in the form of an overall squark condensate  ${|M|^2 \equiv |M_{a}|^2}$, with $a=1,...,N_{f}$ running over the number of flavours $N_{f}$, so that $M_{a} = \sqrt{2 Q^{a} \bar{Q}_{\bar{a}}}$ in terms of the original squarks $(Q^{a},\bar{Q}_{\bar{a}})$ \cite{Achucarro:2006zf}.

Interactions are encoded in a superpotential $\,W\,$ with two contributions:
\begin{equation}
\label{Superpotential}
W(S,T,U,M) =  W_{\textrm{flux}}(S,U) + W_{\textrm{np}}(T,M) \ .
\end{equation}

Let us first discuss the properties of $\,\mathbb{T}^{6}/\mathbb{Z}^{2}_{2}\,$ flux compactifications described by moduli $S$, $T$ and $U$ and perturbative superpotential $W_{\textrm{flux}}(S,U)$.
The latter is induced by RR $\bar{F}_{3}$ and NSNS $\bar{H}_{3}$ background fluxes threading the internal space, giving rise to a GVW-type superpotential \cite{Gukov:1999ya} of polynomial form
\begin{equation}
\label{Wflux}
\begin{array}{lll}
\hspace{-1.4mm}W_{\textrm{flux}}(S,U) &=& \displaystyle \int_{\mathbb{T}^{6}/\mathbb{Z}^{2}_{2}} (\bar{F}_{3} - S \, \bar{H}_{3}) \wedge \Omega(U) \\[4mm]
&=&   ( a_0  - S  \, b_0 ) - 3 \, ( a_1  - S  \, b_1 ) \, U \\[2mm]
&+& 3 \, ( a_2  - S  \, b_2 ) \, U^2 - ( a_3  - S  \, b_3 ) \, U^3 \ , 
\end{array}
\end{equation}
where $\,\Omega(U)\,$ is the $\,\mathbb{T}^{6}/\mathbb{Z}^{2}_{2}\,$ holomorphic three-form. The flux superpotential (\ref{Wflux}) is cubic in the complex structure modulus $U$ and linear in the \mbox{axiodilaton $S$}. The integer flux vectors $\,{\vec{a} \equiv (a_0 \,,\, a_1\,,\, a_2 \,,\, a_3)}\,$ and $\,{\vec{b} \equiv (b_0 \,,\, b_1 \,,\, b_2 \,,\, b_3)}\,$ respectively correspond to $\,\bar{F}_{3}\,$ and $\bar{H}_{3}\,$ background fluxes threading the (orientifold-odd) \mbox{$\mathbb{Z}^{2}_{2}$-invariant} three-cycles (we follow the notation of \cite{Font:2008vd}). Turning on non-trivial background fluxes generically induces a tadpole for the RR potential $\,C_{4}\,$ due to the topological term ${\int \bar{H}_{3} \wedge \bar{F}_{3} \wedge C_{4}}$ in the 10D type IIB action. In the $\mathbb{T}^{6}/\mathbb{Z}^{2}_{2}\,$ case, this translates into the algebraic relation \cite{Font:2008vd}
\begin{equation}
\label{C4_tadpole}
 a_0 \, b_3 - 3 \, a_1 \, b_2 + 3 \, a_2 \, b_1 - a_3 \, b_0 = \tfrac{1}{2} \, N_{\textrm{O3}} - N_{\textrm{D3}} \ ,
\end{equation}
where $N_{\textrm{O3}}=64$ accounts for the O3-planes located at the fixed points of the orientifold action and $N_{\textrm{D3}}$ denotes the number of D3-branes needed to cancel the flux-induced tadpole. Note that there is an upper bound for the flux-induced tadpole (\ref{C4_tadpole}) given by the total O3-plane charge. In contrast, there is no flux-induced tadpole for the O7/D7 sources \cite{Aldazabal:2006up}, thus setting ${\tfrac{1}{2} N_{\textrm{O7}} = N_{\textrm{D7}}=32}\,$.

The second term in (\ref{Superpotential}) models a simple example of non-perturbative contribution to the superpotential involving the K\"ahler modulus $\,T\,$ and the matter field $M$. Following \cite{Achucarro:2006zf,Haack:2006cy}, we take the standard ADS-like superpotential of the form \cite{Taylor:1982bp,Affleck:1983mk}
\begin{equation}
\label{Wnp}
W_{\textrm{np}}(T,M) = (N_c -N_f) \left( \frac{2}{M^{2N_{f}}} \right)^{\tfrac{1}{N_{c}-N_{f}}} e^{i \tfrac{4 \pi k_{N_c}}{N_{c}-N_{f}} T} \  .
\end{equation}
In string theory, these contributions can be generated \textit{e.g.} by gaugino condensation in a non-Abelian sector of $N_{c}$ D7-branes wrapped on some four-cycles of the internal space, or by D-brane instantons.
However, here we do not seek a full implementation in the $\mathbb{T}^6/\mathbb{Z}^2_{2}$ \mbox{orbifold} \cite{Haack:2006cy}, but rather study the resulting toy model with a bottom-up point of view.
In writing (\ref{Wnp}), we are neglecting corrections to the tree-level gauge kinetic function
\begin{equation}
\label{gauge_kin_f}
f_{N_{c}} = \frac{T}{4\pi} + ... \ ,
\end{equation}
which in a string-theoretic framework would account for flux-dependent and/or curvature corrections possibly depending on complex structure, dilaton and open string moduli \cite{Haack:2006cy}.

Combining fluxes and non-perturbative effects, an \mbox{F-term} scalar potential $\,V_{F}(S,T,U,M)\,$ is obtained from (\ref{Kahler}) and (\ref{Superpotential}) via the standard $\,\mathcal{N}=1\,$ formula\footnote{In this letter we are using (reduced) Planck units where ${m_{P}=(8 \, \pi \, G_{\textrm{N}})^{-1/2}= 2.435 \times 10^{18} \textrm{ GeV}}$.}
\begin{equation}
\label{V_N=1}
V_{F}= e^{K} \Big[ K^{\Phi \bar{\Phi} }  D_{\Phi} W \, D_{\bar{\Phi}} \bar{W}  - 3 |W|^2 \Big] \ ,
\end{equation}
where $K^{\Phi \bar{\Phi} }$ is the inverse of the  K\"ahler metric $K_{\Phi \bar{\Phi} }=\partial_{\Phi}\partial_{\bar{\Phi}}K$ and  $D_{\Phi} W=\partial_{\Phi} W + (\partial_{\Phi} K) W$ is the K\"ahler derivative. This time we have collectively denoted all the fields by ${\Phi = (S,T,U,M)}$.  Rewriting the non-perturbative term in (\ref{Wnp}) as $\,W_{\textrm{np}}=|W_{\textrm{np}}| e^{i \varphi_{\textrm{np}}}\,$, one identifies a phase
\begin{equation}
\varphi_{\textrm{np}} = -\frac{2 N_{f}}{N_{c}-N_{f}} \textrm{Arg}(M) + \frac{4 \pi k_{N_c}}{N_{c}-N_{f}} \textrm{Re}(T) \ .
\label{phi_np}
\end{equation}
In terms of this phase $\varphi_{\textrm{np}}$, the potential in (\ref{V_N=1}) can be expressed as
\begin{equation}
\label{VF_np}
V_{F} = h + h_{c} \, \cos\varphi_{\textrm{np}} + h_{s} \, \sin\varphi_{\textrm{np}} \ ,
\end{equation}
where $h(S,U,\textrm{Im}(T),|M| ;  \vec{a},\vec{b})$, $h_{c}(S,U,\textrm{Im}(T),|M| ;  \vec{a},\vec{b})$ and $h_{s}(S,U,\textrm{Im}(T),|M|;a_{1},a_{2},a_{3},\vec{b})$ are lengthy functions depending on the moduli fields and flux parameters we are not displaying here. It is worth emphasising that $h_{s}$ does \textit{not} depend on the flux parameter $a_{0}$, which is the field-independent contribution to (\ref{Wflux}). If setting all but the $a_0$ flux parameter to zero, and \mbox{renaming} it as $W_{0} \equiv a_0\,$, then the superpotential of the STU$|$M-model (\ref{Superpotential}) reduces to that of ACCD~(\ref{W0+Wnp_D-term}).\footnote{Note that, on the contrary, the scalar potential of the STU$|$M-model does not reduce to the one of ACCD due to the presence of the $S$ and $U$ moduli in the K\"ahler potential (\ref{Kahler}).} In the ACCD model, the scalar potential simplifies to $V_{F} = h(\textrm{Im}(T),|M|  ; W_{0}) + h_{c}(\textrm{Im}(T),|M|  ; W_{0})  \cos\varphi_{\textrm{np}}$ and has AdS extrema at $\,\varphi_{\textrm{np}}=  n \, \pi\,$, with $\,n\,$ being an arbitrary integer \cite{Achucarro:2006zf}. This picture is nevertheless modified as long as the $S\,$ and $\,U\,$ moduli enter the flux-induced superpotential (\ref{Wflux}) and $h_{s} \neq 0$ in (\ref{VF_np}). The stabilisation of $\,\varphi_{\textrm{np}}\,$ yields
\begin{equation}
\label{phi_npVEV}
\varphi_{\textrm{np}}= \arctan(h_{s}/h_{c}) + n \, \pi \ ,
\end{equation}
again with $n$ being an arbitrary integer. However, due to the simultaneous presence of the $\cos$ and $\sin$ functions in (\ref{VF_np}), the dS extrema we will present in the next section are only compatible with either even or odd values of $\,n\,$.  

The linear combination of $\textrm{Arg}(M)$ and $\textrm{Re}(T)$ orthogonal to \eqref{phi_np} is associated with a U(1) symmetry of the model, hence it corresponds to a flat direction of the scalar potential.
As such, it is only acceptable if associated with a Goldstone or St\"uckelberg field of a broken gauge symmetry.
This is the case for instance if the matter fields have anomalous charges under some U$(1)_X$, compensated by a shift in the $T$ axion.
In this case, an additional D-term potential is generated to be discussed in sec.~\ref{Sec:D-terms}.
Otherwise, the U(1) symmetry can be explicitly broken by extra perturbative terms in the superpotential, such as a mass term for the matter fields.

Let us now introduce the mass$^{2}$ matrix for the eight real field components $\,\phi^{\alpha}\,$ inside ${\Phi = (S,T,U,M)}$. It is given by
\begin{equation}
\label{m^2_matrix}
[m^2]^{\alpha}{}_{\gamma} = \left. K^{\alpha \beta} \frac{\partial V}{\partial \phi^\beta \partial \phi^\gamma} \right|_{\phi_{0}} \ ,
\end{equation}
where $\,V\,$ is the scalar potential in (\ref{V_N=1}), $\,K^{\alpha \beta}\,$ is the inverse of the K\"ahler metric extracted from the kinetic terms ${L_{\textrm{kin}}=-\tfrac{1}{2}\, K_{\alpha \beta}(\phi) \, \partial\phi^{\alpha} \partial\phi^{\beta}}$ in the real basis and $\phi_{0}\,$ \mbox{denotes} evaluation at a critical point of (\ref{V_N=1}).

We close the section by pointing out a scaling feature of the STU$|$M-models defined by (\ref{Kahler}) and (\ref{Superpotential}). Let us first apply a rescaling of the terms entering the non-perturbative superpotential in (\ref{Wnp}):
\begin{equation}
\label{scaling:N's}
(N_{c},N_{f}) \rightarrow  (\lambda  N_{c} , \lambda  N_{f})
\hspace{2mm} , \hspace{2mm}
T \rightarrow \lambda \, T
\hspace{2mm} , \hspace{2mm}
M \rightarrow \lambda^{-\frac{1}{2}} \, M \ ,
\end{equation}
where $\lambda$ is a scaling parameter ($\lambda N_{c}$ and $\lambda N_{f}$ must \mbox{remain} positive integers). Then, $W_{\textrm{np}}$ changes as
\begin{equation}
\label{scaling:Wnp}
W_{\textrm{np}} \rightarrow \lambda_{\textrm{np}} \, W_{\textrm{np}} \ ,
\end{equation}
with $\,\lambda_{\textrm{np}} \equiv 2^{\frac{1-\lambda}{\lambda (N_{c}-N_{f})}} \lambda^{\frac{N_{c}}{(N_{c}-N_{f})}}$. Transforming now the fluxes and the axiodilaton field as
\begin{equation}
\label{scaling:fluxes}
\vec{a} \rightarrow \lambda_{\textrm{np}} \, \vec{a}
\hspace{4mm} , \hspace{4mm}
\vec{b} \rightarrow \lambda_{\textrm{np}}^{-\beta} \, \vec{b}
\hspace{4mm} , \hspace{4mm}
S \rightarrow \lambda_{\textrm{np}}^{1+\beta} \, S \ ,
\end{equation}
with an arbitrary $\beta \in \mathbb{R}$, then the full superpotential transforms as $\,W \rightarrow \lambda_{\textrm{np}} W\,$ whereas the K\"ahler potential does it as $\,e^{K} \rightarrow \lambda^{-3} \, \lambda_{\textrm{np}}^{-(1+\beta)} \,  e^{K}$. The outcome is an overall scaling of the potential (\ref{V_N=1}) of the form ${V_{F} \rightarrow \lambda^{-3} \, \lambda_{\textrm{np}}^{1-\beta}\, V_{F}}$. Furthermore, in the case of $\beta=1$, the tadpole cancellation condition (\ref{C4_tadpole}) is not affected by the scaling. We will come back to this scaling property in the next sections.

\section{Metastable de Sitter vacuum}
\label{Sec:Examples}

Here we present the main features of the metastable dS vacuum as well as several dS saddle points that we have found in the context of the $\,\mathcal{N}=1\,$ STU$|$M-models specified by the K\"ahler potential (\ref{Kahler}) and superpotential (\ref{Superpotential}) with flux and non-perturbative contributions (\ref{Wflux}) and (\ref{Wnp}). The dS extrema we find are compatible with a choice of parameters of the form
\begin{equation}
\label{parameters_1}
N_{c}=32
\hspace{5mm} , \hspace{5mm}
N_{f}=1
\hspace{5mm} , \hspace{5mm}
k_{N_{c}}=\tfrac{1}{2} \ .
\end{equation}
Recalling the O7/D7 tadpole discussion of the previous section for the toy $\,\mathbb{T}^{6}/\mathbb{Z}^2_{2}\,$ flux model,  it proves natural to take $\,N_{c}\,$ equal to $\,N_{\text{D7}}=32\,$. The number of flavours is set to $N_{f}=1\,$ for simplicity. We also fix ${k_{N_{c}}=1/2}\,$ as in \cite{deCarlos:2007dp}, in order to have a gauge kinetic function $\,{f_{N_{c}}=T /4\pi}\,$ (see (\ref{gauge_kin_f})). Finally, we choose background fluxes
\begin{equation}
\label{fluxes_example_1}
\vec{a} = ( -33 \,,\, 8 \,,\, 10 \,,\, -16)
\hspace{3mm} , \hspace{3mm}
\vec{b} = ( 2 \,,\, 1 \,,\, -7 \,,\, 8) \ ,
\end{equation}
which are enforced to be integers due to the flux quantisation condition. The substitution of the flux parameters (\ref{fluxes_example_1}) into the tadpole cancellation condition (\ref{C4_tadpole}) yields ${N_{\text{D3}}=66}\,$. Therefore, \mbox{D3-branes} are required to cancel the flux-induced tadpole. A similar situation occurs in the flux-scaling scenarios of \cite{Blumenhagen:2015kja,Blumenhagen:2015xpa} where non-geometric fluxes are invoked to stabilise the modulus $T$ in an AdS vacuum. However, this is in contrast to what happens in non-perturbative IIB string scenarios of (closed string) moduli stabilisation in a dS vacuum: O3-planes are typically needed to cancel the flux-induced tadpole \cite{Blaback:2013qza,Marsh:2014nla}, sometimes creating some tension with the upper bound on $\,N_{\text{O3}}\,$ discussed below (\ref{C4_tadpole}).

\begin{table}[t!]
\renewcommand{\arraystretch}{1.5}
\scalebox{0.82}{
\begin{tabular}{!{\vrule width 1.5pt}c!{\vrule width 1pt}c!{\vrule width 1pt}c!{\vrule width 1pt}c!{\vrule width 1pt}c!{\vrule width 1pt}}
\noalign{\hrule height 1.5pt}
                                             &  \sc{minimum} &  \sc{saddle$_1$} &  \sc{saddle$_2$}  &  \sc{saddle$_3$}   \\
\noalign{\hrule height 1pt}
    $S_{0}$           &  $-0.385 + i 1.257$    &  $-0.072 + i 1.452$   &  $0.024 + i 1.537$ &  $-1.195 + i 0.033$ \\
    $U_{0}$           &  $-0.015 + i 0.472$   &  $\phantom{-}0.098 + i 0.510$  &  $0.132 + i 0.596$ &  $-1.477 + i 0.209$  \\
    $T_{0}$           &  $\phantom{-} 0.901 + i 1.518$ &  $\phantom{-} 0.453 + i 1.776$  &  $0.512 + i 3.243$  &  $\phantom{-}17.456 + i 37.146$      \\
    $M_{0}$        &  $0.226$     &  $0.170$  &  $0.109$   &  $0.010$    \\
    $V_{0}$           &  $1.587 \times 10^{-2}$  &  $7.896 \times 10^{-2}$    &  $13.327 \times 10^{-2}$ &  $7.847 \times 10^{-4}$     \\
 \noalign{\hrule height 1.5pt}
\end{tabular}}
\caption{Moduli vev's and cosmological constants for the set of dS critical points of the F-term scalar potential (\ref{V_N=1}). The parameters are set to the values in (\ref{parameters_1}) and (\ref{fluxes_example_1}). These points are compatible with even values of $\,n\,$  in (\ref{phi_npVEV}) (the case $\,n=0\,$ is displayed), except \textsc{saddle$_3$} which requires odd values of $\,n\,$ (the case $\,n=1\,$ is displayed).}
\label{tab:dS}
\end{table}

For our choice of parameters, the F-term scalar potential (\ref{V_N=1}) possesses several dS critical points summarised in TABLE~\ref{tab:dS}. One of these points has 
positive mass eigenvalues, with the exception of the flat direction 
discussed in the previous section.
We anticipate that it can be either eaten up by the massive vector of an anomalous U$(1)_X$ or lifted by an explicit mass term without spoiling the other properties of the vacuum.
Thus, we denote the vacuum expectation value (vev) of the fields at this extremum as $\Phi_{\textrm{min}}$. Note that the physical requirements for a perturbative string coupling constant $g^{-1}_{s}=\textrm{Im}(S_{\textrm{min}}) \gtrsim 1\,$ and for neglecting $\alpha'$-corrections, $\textrm{Im}(T_{\textrm{min}}) \gtrsim 1\,$, are satisfied, although larger values would improve the validity of the supergravity approximation.\footnote{A sporadic search in the parameter space shows that larger values can be achieved, although we have not performed a systematic scanning of integer-valued flux parameters. Alternatively, one could invoke the scaling transformations in (\ref{scaling:N's}) and (\ref{scaling:fluxes}) to get larger values of $\textrm{Im}(T_{\textrm{min}})$ and $\textrm{Im}(S_{\textrm{min}})$. However, the fluxes remain no longer integer-valued after the scaling transformation.}
The computation of the eight eigenvalues of the mass$^{2}$ matrix in (\ref{m^2_matrix}) gives
\begin{equation}
\label{m^2_min_example}
\begin{array}{rrrrrrrrrr}
m^2_{\textrm{min}} &=& 68.475 &,& 28.239 &,&  12.262 &,& 5.528 & , \\[2mm]
         &&  3.959 &,& 2.214 &,& 1.423 &,& 0 & ,
\end{array}
\end{equation}
where the heaviest mode is mostly aligned ($\sim 90 \%$) with $|M|$. As already anticipated, there is a flat direction associated to the axion combination orthogonal to (\ref{phi_np}), which we will remove later on. Note also that there is no large hierarchy amongst the  moduli masses as a result of the single-step stabilisation process. Therefore, the KKLT-like sequential stabilisation is not a justified approximation in this scenario. This is so also for the pattern of supersymmetry breaking. Evaluating the \mbox{F-terms} $F_{\Phi}=e^{K/2}|D_{\Phi}W|\,$ at the minimum, one finds
\begin{equation}
\label{F-terms}
\begin{array}{llll}
F_{S}= 0.586  & \hspace{5mm} , \hspace{5mm}  & \,\,\,\,\,\,\,\, F_{U}= 2.840 & , \\[2mm]
F_{T}= 2.311  & \hspace{5mm} , \hspace{5mm}  & \,\,\,\,\,\,\, F_{M}= 1.488 & ,
\end{array}
\end{equation}
and SUSY is broken in both the $\,(S,U)\,$ and $\,(T,M)\,$ sectors in a sizable way. More concretely, $\,F_{U}\sim F_{T}\,$ clashes with the decoupling between complex structure and K\"ahler moduli assumed in KKLT-like scenarios.
Also, when evaluated at the minimum, there is a dynamically \mbox{generated} gravitino mass $\,m_{3/2}= (4 \pi)^{-\frac{1}{2}}  e^{K/2} |W|  \, m_{P}\,$ of the same order as the moduli masses in (\ref{m^2_min_example}). 
In other words, $\,m_{3/2}\sim m_{\textrm{moduli}}\,$, rendering the model unsuitable for \mbox{phenomenology}.

Assuming that the flat direction is removed by one of the mechanisms to be discussed shortly, we could ask whether one of the other dS points with tachyonic directions (labeled \mbox{\textsc{saddle$_{1,2,3}$}} in TABLE~\ref{tab:dS}) could accommodate for a slow-roll scenario towards the metastable vacuum. This automatically excludes \textsc{saddle$_3$} from the analysis as it has a lower potential energy.\footnote{Moreover, \textsc{saddle$_3$} becomes non-perturbative in the string coupling constant $g_{s}=1/\textrm{Im}(S_{0})\,$, although \mbox{$\alpha$'-corrections} are highly suppressed as $\textrm{Im}(T_{0}) \gg 1$.} 
The remaining \textsc{saddle$_{1,2}$} points are monotonically connected to the minimum, but the instabilities are too large to accommodate for slow-roll inflation. 
The presence of large tachyon masses, which is usually referred to as $\eta$-problem in supergravity, is a common feature in single-step moduli stabilisation scenarios where separation of scales does generically not occur and where the integer nature of the flux parameters does not allow for a continuous tuning.

Finally, let us consider the possibility of lifting the flat direction in (\ref{m^2_min_example}) by explicitly introducing a mass term for the squarks \cite{Intriligator:2006dd}.
This amounts to adding a term of the form $\mu \,  M^2$ to the superpotential in (\ref{Superpotential}) with a constant mass parameter $\mu$. 
For small positive values of $\mu$, the flat direction is lifted without destabilising the original vacuum at $\mu=0$.
The value of the cosmological constant $V_{\textrm{min}}\,$ decreases as long as $\,\mu\,$ increases, thus down-lifting the de Sitter minimum to either Minkowski, at certain critical value $\,\mu_{0}=0.0735053\,$, or to Anti-de-Sitter if $\mu>\mu_{0}$, preserving stability. Similar scenarios of spontaneous SUSY breaking based on the ISS \mbox{(or Polonyi)} model in combination with KKLT-like models of K\"ahler modulus stabilisation have been investigated in the literature (see \cite{Abe:2006xp} and references therein). Moreover, hidden sector matter interactions were shown to succeed in (\mbox{F-term}) uplifting the AdS vacuum of KKLT to a dS one \cite{Lebedev:2006qq,Abe:2006xp,Kallosh:2006dv}.

\section{Anomalous \boldmath $\textrm{U}(1)_{X}$ and D-terms}
\label{Sec:D-terms}

The solutions discussed in the previous section are obtained from a pure F-term scalar potential.
However, they are also compatible with extra D-term contributions arising when the matter fields carry anomalous charges with respect to some U$(1)_X$ gauge symmetry, rendered consistent by a Green-Schwarz mechanism involving the $T$ axion \cite{Binetruy:1996uv,Burgess:2003ic,Villadoro:2005yq}.
In a string setting, this U$(1)_X$ may arise for instance from the gauge symmetries of stacks of \mbox{D7-branes} \cite{Achucarro:2006zf,Haack:2006cy}.
Contrary to the situation depicted \textit{e.g.} in \cite{Achucarro:2006zf}, these extra contributions are not necessary ingredients for the existence of a de Sitter vacuum, although they are compatible with it.
Rather, they provide a natural mechanism to remove the flat direction of the pure F-term potential, as it is eaten up by the massive U$(1)_X$ vector.

Consider the squark $Q$ and anti-squark $\bar{Q}$ to have charges $q$ and $\bar{q}$ under a $\textrm{U}(1)_{X}$ transformation with parameter $\epsilon$. 
If it is anomalous, namely $\,q + \bar{q}\neq0\,$, then a non-vanishing \mbox{D-term} is generated which takes the form
\begin{equation}
\label{VD}
V_{D}=\frac{i \pi}{2 \, k_X\,  (T-\bar{T})} \left( (q+\bar{q}) |M|^2 - \frac{3 \, i  \, \delta_{\textrm{GS}}}{T-\bar{T}}\right)^2 \ ,
\end{equation}
where $\delta_{\textrm{GS}}$ denotes the $\textrm{U}(1)_{X}$ charge of the $T$ \mbox{modulus}, transforming as $T \rightarrow T + q(T) \epsilon$, via the Green-Schwarz anomaly cancellation mechanism. There are $\textrm{SU}(N_{c})^{2}\times\textrm{U}(1)_{X}\,$ and $\,\textrm{U}(1)^{3}_{X}\,$ gauge anomalies arising from triangular Feynman diagrams, whose cancellation sets $\,{q(T)=-\delta_{\textrm{GS}}/2}\,$ with \cite{Binetruy:1996uv,Achucarro:2006zf}
\begin{equation}
\delta_{\textrm{GS}} = -\frac{N_{f}}{2 \, \pi \, k_{N_{c}}} (q+\bar{q})
\hspace{2mm} , \hspace{2mm}
k_{X} = \frac{2}{3} \, N_{c} \, k_{N_{c}} \left( \frac{q^3 + \bar{q}^3}{q+\bar{q}} \right) \ .
\end{equation}
Therefore, the D-term contribution (\ref{VD}) to the scalar potential $\,V=V_{F}+V_{D}\,$ is totally encoded into the choice of anomalous $\textrm{U}(1)_{X}$ charges $(q,\bar{q})$. 

\begin{table}[t!]
\renewcommand{\arraystretch}{1.5}
\begin{tabular}{!{\vrule width 1.5pt}c!{\vrule width 1pt}c!{\vrule width 1pt}c!{\vrule width 1pt}c!{\vrule width 1pt}}
\noalign{\hrule height 1.5pt}
    $(q,\bar{q})$          &  $(3,-2)$ &  $(4,7)$  &  $(1,1)$  \\
\noalign{\hrule height 1pt}
    $S_{\textrm{min}}$           &  $-0.385 + i 1.257$ &  $-0.376 + i 1.253$  &  $-0.374 + i 1.252$    \\
    $U_{\textrm{min}}$           &  $-0.015 + i 0.472$ &  $-0.016 + i 0.475$ &  $-0.016 + i 0.476$    \\
    $T_{\textrm{min}}$           &  $\phantom{-} 0.901 + i 1.519$  &  $\phantom{-} 0.909 + i 1.556$  &  $\phantom{-} 0.911 + i 1.565$   \\
    $|M_{\textrm{min}}|$        &  $0.226$   &  $0.223$   &  $0.222$     \\
    $V_{\textrm{min}}$            &  $1.621 \times 10^{-2}$ &  $3.631 \times 10^{-2}$  &  $4.066 \times 10^{-2}$      \\
    $(V_{D})_{\textrm{min}}$   &  $3.410 \times 10^{-4}$  &  $1.967 \times 10^{-2}$   &  $2.364 \times 10^{-2}$  \\
 \noalign{\hrule height 1.5pt}
\end{tabular}
\caption{Moduli vev's, cosmological constants and \mbox{D-term} contributions at the minimum of the potential for different assignments of the $\textrm{U}(1)_{X}$ charges $(q,\bar{q})$. The rest of the parameters has been set to their values in (\ref{parameters_1}) and (\ref{fluxes_example_1}).}
\label{tab:D-terms}
\end{table}

We have investigated how the process of moduli stabilisation in a metastable dS vacuum described in sec.~\ref{Sec:Examples} is affected by the D-term. We stress again that in the original ACCD work \cite{Achucarro:2006zf}, which did not include the $S$ and $U$ moduli as dynamical fields, the presence of a non-vanishing D-term was a necessary ingredient in order to uplift an (F-term) AdS vacuum to a dS one. In our setup, the stabilisation in a dS vacuum proceeds directly due to the F-term interplay between the STU moduli and the matter fields, and the effect of turning on a D-term is just a small shift on the moduli vev's and an uplift of the cosmological constant whose size depends on the charges $(q,\bar{q})$. The mass spectrum at the minimum gets slightly modified accordingly and the flat direction is no longer a physical direction. Some numerical results for different charge assignments are collected in TABLE~\ref{tab:D-terms}.

\section{Discussion}
\label{Sec:Discussion}

Following a bottom-up approach, we have investigated classes of $\mathcal{N}=1$ effective supergravities inspired by type~IIB orientifold constructions,  which include non-trivial background fluxes and non-perturbative effects. 
In a string theory setting, the latter can be induced by gaugino condensation in a non-Abelian sector of D7-branes and introduce a coupling between closed string modes and matter fields associated with open string degrees of freedom.
We have found that a purely F-term scalar potential suffices to stabilise all the moduli in a metastable de Sitter vacuum.
The mass spectrum contains a flat direction which is either eaten up by the massive vector of an anomalous U$(1)_X$, or can be lifted by an explicit mass term.
Remarkably, the stabilisation of the moduli in a dS vacuum occurs as a single-step process, thus avoiding the need for some separate mechanism to lift the cosmological constant to positive values.

\begin{figure}[t!]
\centering
\includegraphics[width=87mm]{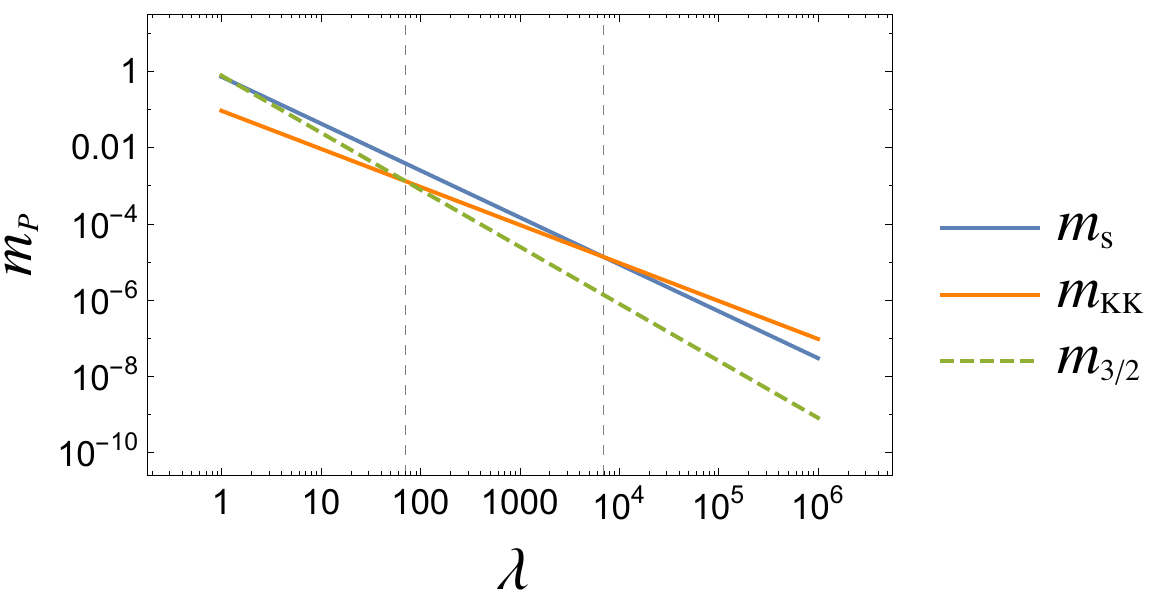}
\caption{Plot of the running of the string ($m_{\textrm{s}}$), KK ($m_{\textrm{KK}}$) and gravitino ($m_{3/2}$) mass scales with the scaling parameter $\lambda$ at the metastable dS vacuum. There is a $\lambda$-interval having $m_{\textrm{s}}  \gtrsim m_{\textrm{KK}} \sim  m_{3/2}$ and $m_{\textrm{s}}  \sim m_{\textrm{KK}} \gtrsim  m_{3/2}$ as boundaries (vertical dashed lines), in which the right order of scales ${m_{\textrm{s}}  \gtrsim m_{\textrm{KK}} \gtrsim  m_{3/2}}$ is achieved. We have set $\beta=1$ in (\ref{scaling:fluxes}) so that $N_{\textrm{D3}}=66$ does not scale with $\lambda$.}
\label{Fig:scales} 
\end{figure}

Some previous results on single-step de Sitter vacua do not seem to achieve a satisfactory hierarchy between the induced string and Kaluza-Klein (KK) scales on one side  and the masses of the supergravity fields and the supersymmetry breaking scale on the other. Taking as working definitions for the string ($m_{\textrm{s}}$) and KK ($m_{\textrm{KK}}$) scales \cite{Blumenhagen:2009gk}
\begin{equation}
\label{KK and string scales}
\begin{array}{llll}
m_{\textrm{s}} &=& \pi^{\frac{1}{2}} \, \textrm{Im}(S_{\textrm{min}})^{-\frac{1}{4}}(2 \, \textrm{Im}(T_{\textrm{min}}))^{-\frac{3}{4}} \, m_{P} \ ,\\[2mm]
m_{\textrm{KK}} &=& (4 \pi)^{-\frac{1}{2}} (2 \, \textrm{Im}(T_{\textrm{min}}))^{-1} \, m_{P}\ ,
\end{array}
\end{equation}
a direct computation shows for instance that in \cite{Blaback:2013qza,Marsh:2014nla} some of the supergravity fields have masses larger than the KK and string scales. This would invalidate the effective supergravity approximation which additionally requires the moduli VEVs to remain in the perturbative regime, namely, $\textrm{Im}(S_{\textrm{min}})\gtrsim 1$ and $\textrm{Im}(T_{\textrm{min}}) \gtrsim 1$. Interestingly, the correct hierarchy of scales can be easily obtained in certain non-geometric flux compactification scenarios \cite{Blumenhagen:2015kja,Blumenhagen:2015xpa} which feature enough scaling properties to control the values of $W$ (and thus $V$), $\,\textrm{Im}(T)\,$ and $\,\textrm{Im}(S)$ independently at the dS  minimum, thus giving parametric control over the relevant mass ratios. This is achieved at the cost of introducing hierarchies amongst the fluxes.
A situation similar to \cite{Blaback:2013qza,Marsh:2014nla} occurs in the solution of sec.~\ref{Sec:Examples}.
Although we do not have as much scaling freedom as \cite{Blumenhagen:2015kja,Blumenhagen:2015xpa}, we still find that the \mbox{$\lambda$-scaling} in (\ref{scaling:N's}) and (\ref{scaling:fluxes}) is sufficient to achieve the correct hierarchy of scales in a certain window in the range of $\lambda$ \mbox{(see FIG.~\ref{Fig:scales})}. As a result, large hierarchies amongst the fluxes are induced. 
Whether the above findings indicate a fundamental tension between a consistent hierarchy of scales and single-step de Sitter vacua in the supergravity regime is certainly an interesting question which should be further investigated. 
It would then be interesting to carry out a systematic scanning of the parameter space to search for dS vacua that yield the correct hierarchy of scales without introducing hierarchies amongst the fluxes. We postpone this thorough analysis to future work in the context of a string theoretic realisation of these scenarios.

Despite recent progress in deriving semi-analytic methods to hunt for (single-step) de Sitter solutions in the framework of $\mathcal{N}=1$ effective supergravities including \mbox{(non-)geometric} fluxes \cite{deCarlos:2009fq,deCarlos:2009qm,Danielsson:2012by,Blaback:2013ht}, \mbox{(non-)perturbative} effects \cite{Blaback:2013qza,Danielsson:2013rza,Kallosh:2014oja}, etc., it still remains a big challenge to understand what are the physical principles leading to a stable de Sitter vacuum in string theory.
The scenario presented here should be viewed as a proof of concept that metastable dS vacua can be obtained from F-term potentials, including matter fields and as a single-step process, in regions of parameter space that take into account flux quantisation and are usually computationally difficult to probe.
This clearly opens the possibility of looking for a more thorough string realisation of the scenario we have described. This would also allow us to gain more control on certain approximations. For instance, the tree-level gauge-kinetic function in (\ref{gauge_kin_f}) could be corrected by extra flux-dependent terms, thus affecting the non-perturbative superpotential \cite{Witten:1996bn,Haack:2006cy,Blaback:2013qza}. Also the backreaction of the (large) fluxes/D-branes on the geometry or modifications of the K\"ahler potential for the matter fields of the form $K \sim N_{f}|M|^2/\Phi$ ($\Phi$ being some combination of closed string moduli), should be appropriately taken into account \cite{Ibanez:1998rf,Dudas:2006gr}. We hope to come back  to these problems in the future.

\noindent\textbf{Acknowledgements:} We thank Ana Ach\'ucarro and Diederik Roest for discussions and valuable comments on the manuscript, and Gerardus Wittems for arming us with the patience to hunt the de Sitter extrema presented in this letter. 
The work of AG and GI is supported by the ERC Advanced Grant no. 246974, {``Supersymmetry: a window to non-perturbative physics''}.

\bibliography{references}

\begin{thebibliography}{42}
\expandafter\ifx\csname natexlab\endcsname\relax\def\natexlab#1{#1}\fi
\expandafter\ifx\csname bibnamefont\endcsname\relax
  \def\bibnamefont#1{#1}\fi
\expandafter\ifx\csname bibfnamefont\endcsname\relax
  \def\bibfnamefont#1{#1}\fi
\expandafter\ifx\csname citenamefont\endcsname\relax
  \def\citenamefont#1{#1}\fi
\expandafter\ifx\csname url\endcsname\relax
  \def\url#1{\texttt{#1}}\fi
\expandafter\ifx\csname urlprefix\endcsname\relax\def\urlprefix{URL }\fi
\providecommand{\bibinfo}[2]{#2}
\providecommand{\eprint}[2][]{\url{#2}}

\bibitem[{\citenamefont{Kachru et~al.}(2003)\citenamefont{Kachru, Kallosh,
  Linde, and Trivedi}}]{Kachru:2003aw}
\bibinfo{author}{\bibfnamefont{S.}~\bibnamefont{Kachru}},
  \bibinfo{author}{\bibfnamefont{R.}~\bibnamefont{Kallosh}},
  \bibinfo{author}{\bibfnamefont{A.~D.} \bibnamefont{Linde}}, \bibnamefont{and}
  \bibinfo{author}{\bibfnamefont{S.~P.} \bibnamefont{Trivedi}},
  \bibinfo{journal}{Phys. Rev.} \textbf{\bibinfo{volume}{D68}},
  \bibinfo{pages}{046005} (\bibinfo{year}{2003}), \eprint{hep-th/0301240}.

\bibitem[{\citenamefont{Giddings et~al.}(2002)\citenamefont{Giddings, Kachru,
  and Polchinski}}]{Giddings:2001yu}
\bibinfo{author}{\bibfnamefont{S.~B.} \bibnamefont{Giddings}},
  \bibinfo{author}{\bibfnamefont{S.}~\bibnamefont{Kachru}}, \bibnamefont{and}
  \bibinfo{author}{\bibfnamefont{J.}~\bibnamefont{Polchinski}},
  \bibinfo{journal}{Phys. Rev.} \textbf{\bibinfo{volume}{D66}},
  \bibinfo{pages}{106006} (\bibinfo{year}{2002}), \eprint{hep-th/0105097}.

\bibitem[{\citenamefont{Derendinger et~al.}(1985)\citenamefont{Derendinger,
  Ibanez, and Nilles}}]{Derendinger:1985kk}
\bibinfo{author}{\bibfnamefont{J.~P.} \bibnamefont{Derendinger}},
  \bibinfo{author}{\bibfnamefont{L.~E.} \bibnamefont{Ibanez}},
  \bibnamefont{and} \bibinfo{author}{\bibfnamefont{H.~P.}
  \bibnamefont{Nilles}}, \bibinfo{journal}{Phys. Lett.}
  \textbf{\bibinfo{volume}{B155}}, \bibinfo{pages}{65} (\bibinfo{year}{1985}).

\bibitem[{\citenamefont{Dine et~al.}(1985)\citenamefont{Dine, Rohm, Seiberg,
  and Witten}}]{Dine:1985rz}
\bibinfo{author}{\bibfnamefont{M.}~\bibnamefont{Dine}},
  \bibinfo{author}{\bibfnamefont{R.}~\bibnamefont{Rohm}},
  \bibinfo{author}{\bibfnamefont{N.}~\bibnamefont{Seiberg}}, \bibnamefont{and}
  \bibinfo{author}{\bibfnamefont{E.}~\bibnamefont{Witten}},
  \bibinfo{journal}{Phys. Lett.} \textbf{\bibinfo{volume}{B156}},
  \bibinfo{pages}{55} (\bibinfo{year}{1985}).

\bibitem[{\citenamefont{Burgess et~al.}(2003)\citenamefont{Burgess, Kallosh,
  and Quevedo}}]{Burgess:2003ic}
\bibinfo{author}{\bibfnamefont{C.~P.} \bibnamefont{Burgess}},
  \bibinfo{author}{\bibfnamefont{R.}~\bibnamefont{Kallosh}}, \bibnamefont{and}
  \bibinfo{author}{\bibfnamefont{F.}~\bibnamefont{Quevedo}},
  \bibinfo{journal}{JHEP} \textbf{\bibinfo{volume}{10}}, \bibinfo{pages}{056}
  (\bibinfo{year}{2003}), \eprint{hep-th/0309187}.

\bibitem[{\citenamefont{Villadoro and Zwirner}(2005)}]{Villadoro:2005yq}
\bibinfo{author}{\bibfnamefont{G.}~\bibnamefont{Villadoro}} \bibnamefont{and}
  \bibinfo{author}{\bibfnamefont{F.}~\bibnamefont{Zwirner}},
  \bibinfo{journal}{Phys. Rev. Lett.} \textbf{\bibinfo{volume}{95}},
  \bibinfo{pages}{231602} (\bibinfo{year}{2005}), \eprint{hep-th/0508167}.

\bibitem[{\citenamefont{Balasubramanian and
  Berglund}(2004)}]{Balasubramanian:2004uy}
\bibinfo{author}{\bibfnamefont{V.}~\bibnamefont{Balasubramanian}}
  \bibnamefont{and} \bibinfo{author}{\bibfnamefont{P.}~\bibnamefont{Berglund}},
  \bibinfo{journal}{JHEP} \textbf{\bibinfo{volume}{11}}, \bibinfo{pages}{085}
  (\bibinfo{year}{2004}), \eprint{hep-th/0408054}.

\bibitem[{\citenamefont{von Gersdorff and
  Hebecker}(2005)}]{vonGersdorff:2005bf}
\bibinfo{author}{\bibfnamefont{G.}~\bibnamefont{von Gersdorff}}
  \bibnamefont{and} \bibinfo{author}{\bibfnamefont{A.}~\bibnamefont{Hebecker}},
  \bibinfo{journal}{Phys. Lett.} \textbf{\bibinfo{volume}{B624}},
  \bibinfo{pages}{270} (\bibinfo{year}{2005}), \eprint{hep-th/0507131}.

\bibitem[{\citenamefont{Berg et~al.}(2006)\citenamefont{Berg, Haack, and
  Kors}}]{Berg:2005yu}
\bibinfo{author}{\bibfnamefont{M.}~\bibnamefont{Berg}},
  \bibinfo{author}{\bibfnamefont{M.}~\bibnamefont{Haack}}, \bibnamefont{and}
  \bibinfo{author}{\bibfnamefont{B.}~\bibnamefont{Kors}},
  \bibinfo{journal}{Phys. Rev. Lett.} \textbf{\bibinfo{volume}{96}},
  \bibinfo{pages}{021601} (\bibinfo{year}{2006}), \eprint{hep-th/0508171}.

\bibitem[{\citenamefont{Parameswaran and Westphal}(2006)}]{Parameswaran:2006jh}
\bibinfo{author}{\bibfnamefont{S.~L.} \bibnamefont{Parameswaran}}
  \bibnamefont{and} \bibinfo{author}{\bibfnamefont{A.}~\bibnamefont{Westphal}},
  \bibinfo{journal}{JHEP} \textbf{\bibinfo{volume}{10}}, \bibinfo{pages}{079}
  (\bibinfo{year}{2006}), \eprint{hep-th/0602253}.

\bibitem[{\citenamefont{Westphal}(2007)}]{Westphal:2006tn}
\bibinfo{author}{\bibfnamefont{A.}~\bibnamefont{Westphal}},
  \bibinfo{journal}{JHEP} \textbf{\bibinfo{volume}{03}}, \bibinfo{pages}{102}
  (\bibinfo{year}{2007}), \eprint{hep-th/0611332}.

\bibitem[{\citenamefont{Rummel and Westphal}(2012)}]{Rummel:2011cd}
\bibinfo{author}{\bibfnamefont{M.}~\bibnamefont{Rummel}} \bibnamefont{and}
  \bibinfo{author}{\bibfnamefont{A.}~\bibnamefont{Westphal}},
  \bibinfo{journal}{JHEP} \textbf{\bibinfo{volume}{01}}, \bibinfo{pages}{020}
  (\bibinfo{year}{2012}), \eprint{1107.2115}.

\bibitem[{\citenamefont{Achucarro et~al.}(2006)\citenamefont{Achucarro,
  de~Carlos, Casas, and Doplicher}}]{Achucarro:2006zf}
\bibinfo{author}{\bibfnamefont{A.}~\bibnamefont{Achucarro}},
  \bibinfo{author}{\bibfnamefont{B.}~\bibnamefont{de~Carlos}},
  \bibinfo{author}{\bibfnamefont{J.~A.} \bibnamefont{Casas}}, \bibnamefont{and}
  \bibinfo{author}{\bibfnamefont{L.}~\bibnamefont{Doplicher}},
  \bibinfo{journal}{JHEP} \textbf{\bibinfo{volume}{06}}, \bibinfo{pages}{014}
  (\bibinfo{year}{2006}), \eprint{hep-th/0601190}.

\bibitem[{\citenamefont{Haack et~al.}(2007)\citenamefont{Haack, Krefl, Lust,
  Van~Proeyen, and Zagermann}}]{Haack:2006cy}
\bibinfo{author}{\bibfnamefont{M.}~\bibnamefont{Haack}},
  \bibinfo{author}{\bibfnamefont{D.}~\bibnamefont{Krefl}},
  \bibinfo{author}{\bibfnamefont{D.}~\bibnamefont{Lust}},
  \bibinfo{author}{\bibfnamefont{A.}~\bibnamefont{Van~Proeyen}},
  \bibnamefont{and}
  \bibinfo{author}{\bibfnamefont{M.}~\bibnamefont{Zagermann}},
  \bibinfo{journal}{JHEP} \textbf{\bibinfo{volume}{01}}, \bibinfo{pages}{078}
  (\bibinfo{year}{2007}), \eprint{hep-th/0609211}.

\bibitem[{\citenamefont{Green and Schwarz}(1984)}]{Green:1984sg}
\bibinfo{author}{\bibfnamefont{M.~B.} \bibnamefont{Green}} \bibnamefont{and}
  \bibinfo{author}{\bibfnamefont{J.~H.} \bibnamefont{Schwarz}},
  \bibinfo{journal}{Phys. Lett.} \textbf{\bibinfo{volume}{B149}},
  \bibinfo{pages}{117} (\bibinfo{year}{1984}).

\bibitem[{\citenamefont{Blåbäck et~al.}(2014)\citenamefont{Blåbäck, Roest,
  and Zavala}}]{Blaback:2013qza}
\bibinfo{author}{\bibfnamefont{J.}~\bibnamefont{Blåbäck}},
  \bibinfo{author}{\bibfnamefont{D.}~\bibnamefont{Roest}}, \bibnamefont{and}
  \bibinfo{author}{\bibfnamefont{I.}~\bibnamefont{Zavala}},
  \bibinfo{journal}{Phys. Rev.} \textbf{\bibinfo{volume}{D90}},
  \bibinfo{pages}{024065} (\bibinfo{year}{2014}), \eprint{1312.5328}.

\bibitem[{\citenamefont{Kallosh et~al.}(2014)\citenamefont{Kallosh, Linde,
  Vercnocke, and Wrase}}]{Kallosh:2014oja}
\bibinfo{author}{\bibfnamefont{R.}~\bibnamefont{Kallosh}},
  \bibinfo{author}{\bibfnamefont{A.}~\bibnamefont{Linde}},
  \bibinfo{author}{\bibfnamefont{B.}~\bibnamefont{Vercnocke}},
  \bibnamefont{and} \bibinfo{author}{\bibfnamefont{T.}~\bibnamefont{Wrase}},
  \bibinfo{journal}{JHEP} \textbf{\bibinfo{volume}{10}}, \bibinfo{pages}{11}
  (\bibinfo{year}{2014}), \eprint{1406.4866}.

\bibitem[{\citenamefont{Marsh et~al.}(2015)\citenamefont{Marsh, Vercnocke, and
  Wrase}}]{Marsh:2014nla}
\bibinfo{author}{\bibfnamefont{M.~C.~D.} \bibnamefont{Marsh}},
  \bibinfo{author}{\bibfnamefont{B.}~\bibnamefont{Vercnocke}},
  \bibnamefont{and} \bibinfo{author}{\bibfnamefont{T.}~\bibnamefont{Wrase}},
  \bibinfo{journal}{JHEP} \textbf{\bibinfo{volume}{05}}, \bibinfo{pages}{081}
  (\bibinfo{year}{2015}), \eprint{1411.6625}.

\bibitem[{\citenamefont{Witten}(1996)}]{Witten:1996bn}
\bibinfo{author}{\bibfnamefont{E.}~\bibnamefont{Witten}},
  \bibinfo{journal}{Nucl. Phys.} \textbf{\bibinfo{volume}{B474}},
  \bibinfo{pages}{343} (\bibinfo{year}{1996}), \eprint{hep-th/9604030}.

\bibitem[{\citenamefont{de~Alwis}(2005)}]{deAlwis:2005tf}
\bibinfo{author}{\bibfnamefont{S.~P.} \bibnamefont{de~Alwis}},
  \bibinfo{journal}{Phys. Lett.} \textbf{\bibinfo{volume}{B626}},
  \bibinfo{pages}{223} (\bibinfo{year}{2005}), \eprint{hep-th/0506266}.

\bibitem[{\citenamefont{Derendinger et~al.}(2005)\citenamefont{Derendinger,
  Kounnas, Petropoulos, and Zwirner}}]{Derendinger:2004jn}
\bibinfo{author}{\bibfnamefont{J.-P.} \bibnamefont{Derendinger}},
  \bibinfo{author}{\bibfnamefont{C.}~\bibnamefont{Kounnas}},
  \bibinfo{author}{\bibfnamefont{P.~M.} \bibnamefont{Petropoulos}},
  \bibnamefont{and} \bibinfo{author}{\bibfnamefont{F.}~\bibnamefont{Zwirner}},
  \bibinfo{journal}{Nucl.Phys.} \textbf{\bibinfo{volume}{B715}},
  \bibinfo{pages}{211} (\bibinfo{year}{2005}), \eprint{hep-th/0411276}.

\bibitem[{\citenamefont{Lebedev et~al.}(2006)\citenamefont{Lebedev, Nilles, and
  Ratz}}]{Lebedev:2006qq}
\bibinfo{author}{\bibfnamefont{O.}~\bibnamefont{Lebedev}},
  \bibinfo{author}{\bibfnamefont{H.~P.} \bibnamefont{Nilles}},
  \bibnamefont{and} \bibinfo{author}{\bibfnamefont{M.}~\bibnamefont{Ratz}},
  \bibinfo{journal}{Phys. Lett.} \textbf{\bibinfo{volume}{B636}},
  \bibinfo{pages}{126} (\bibinfo{year}{2006}), \eprint{hep-th/0603047}.

\bibitem[{\citenamefont{Ibanez et~al.}(1999)\citenamefont{Ibanez, Munoz, and
  Rigolin}}]{Ibanez:1998rf}
\bibinfo{author}{\bibfnamefont{L.~E.} \bibnamefont{Ibanez}},
  \bibinfo{author}{\bibfnamefont{C.}~\bibnamefont{Munoz}}, \bibnamefont{and}
  \bibinfo{author}{\bibfnamefont{S.}~\bibnamefont{Rigolin}},
  \bibinfo{journal}{Nucl. Phys.} \textbf{\bibinfo{volume}{B553}},
  \bibinfo{pages}{43} (\bibinfo{year}{1999}), \eprint{hep-ph/9812397}.

\bibitem[{\citenamefont{Gukov et~al.}(2000)\citenamefont{Gukov, Vafa, and
  Witten}}]{Gukov:1999ya}
\bibinfo{author}{\bibfnamefont{S.}~\bibnamefont{Gukov}},
  \bibinfo{author}{\bibfnamefont{C.}~\bibnamefont{Vafa}}, \bibnamefont{and}
  \bibinfo{author}{\bibfnamefont{E.}~\bibnamefont{Witten}},
  \bibinfo{journal}{Nucl. Phys.} \textbf{\bibinfo{volume}{B584}},
  \bibinfo{pages}{69} (\bibinfo{year}{2000}), \bibinfo{note}{[Erratum: Nucl.
  Phys.B608,477(2001)]}, \eprint{hep-th/9906070}.

\bibitem[{\citenamefont{Font et~al.}(2008)\citenamefont{Font, Guarino, and
  Moreno}}]{Font:2008vd}
\bibinfo{author}{\bibfnamefont{A.}~\bibnamefont{Font}},
  \bibinfo{author}{\bibfnamefont{A.}~\bibnamefont{Guarino}}, \bibnamefont{and}
  \bibinfo{author}{\bibfnamefont{J.~M.} \bibnamefont{Moreno}},
  \bibinfo{journal}{JHEP} \textbf{\bibinfo{volume}{12}}, \bibinfo{pages}{050}
  (\bibinfo{year}{2008}), \eprint{0809.3748}.

\bibitem[{\citenamefont{Aldazabal et~al.}(2006)\citenamefont{Aldazabal, Camara,
  Font, and Ibanez}}]{Aldazabal:2006up}
\bibinfo{author}{\bibfnamefont{G.}~\bibnamefont{Aldazabal}},
  \bibinfo{author}{\bibfnamefont{P.~G.} \bibnamefont{Camara}},
  \bibinfo{author}{\bibfnamefont{A.}~\bibnamefont{Font}}, \bibnamefont{and}
  \bibinfo{author}{\bibfnamefont{L.}~\bibnamefont{Ibanez}},
  \bibinfo{journal}{JHEP} \textbf{\bibinfo{volume}{0605}}, \bibinfo{pages}{070}
  (\bibinfo{year}{2006}), \eprint{hep-th/0602089}.

\bibitem[{\citenamefont{Taylor et~al.}(1983)\citenamefont{Taylor, Veneziano,
  and Yankielowicz}}]{Taylor:1982bp}
\bibinfo{author}{\bibfnamefont{T.~R.} \bibnamefont{Taylor}},
  \bibinfo{author}{\bibfnamefont{G.}~\bibnamefont{Veneziano}},
  \bibnamefont{and}
  \bibinfo{author}{\bibfnamefont{S.}~\bibnamefont{Yankielowicz}},
  \bibinfo{journal}{Nucl. Phys.} \textbf{\bibinfo{volume}{B218}},
  \bibinfo{pages}{493} (\bibinfo{year}{1983}).

\bibitem[{\citenamefont{Affleck et~al.}(1984)\citenamefont{Affleck, Dine, and
  Seiberg}}]{Affleck:1983mk}
\bibinfo{author}{\bibfnamefont{I.}~\bibnamefont{Affleck}},
  \bibinfo{author}{\bibfnamefont{M.}~\bibnamefont{Dine}}, \bibnamefont{and}
  \bibinfo{author}{\bibfnamefont{N.}~\bibnamefont{Seiberg}},
  \bibinfo{journal}{Nucl. Phys.} \textbf{\bibinfo{volume}{B241}},
  \bibinfo{pages}{493} (\bibinfo{year}{1984}).

\bibitem[{\citenamefont{de~Carlos et~al.}(2007)\citenamefont{de~Carlos, Casas,
  Guarino, Moreno, and Seto}}]{deCarlos:2007dp}
\bibinfo{author}{\bibfnamefont{B.}~\bibnamefont{de~Carlos}},
  \bibinfo{author}{\bibfnamefont{J.~A.} \bibnamefont{Casas}},
  \bibinfo{author}{\bibfnamefont{A.}~\bibnamefont{Guarino}},
  \bibinfo{author}{\bibfnamefont{J.~M.} \bibnamefont{Moreno}},
  \bibnamefont{and} \bibinfo{author}{\bibfnamefont{O.}~\bibnamefont{Seto}},
  \bibinfo{journal}{JCAP} \textbf{\bibinfo{volume}{0705}}, \bibinfo{pages}{002}
  (\bibinfo{year}{2007}), \eprint{hep-th/0702103}.

\bibitem[{\citenamefont{Blumenhagen
  et~al.}(2015{\natexlab{a}})\citenamefont{Blumenhagen, Font, Fuchs,
  Herschmann, Plauschinn, Sekiguchi, and Wolf}}]{Blumenhagen:2015kja}
\bibinfo{author}{\bibfnamefont{R.}~\bibnamefont{Blumenhagen}},
  \bibinfo{author}{\bibfnamefont{A.}~\bibnamefont{Font}},
  \bibinfo{author}{\bibfnamefont{M.}~\bibnamefont{Fuchs}},
  \bibinfo{author}{\bibfnamefont{D.}~\bibnamefont{Herschmann}},
  \bibinfo{author}{\bibfnamefont{E.}~\bibnamefont{Plauschinn}},
  \bibinfo{author}{\bibfnamefont{Y.}~\bibnamefont{Sekiguchi}},
  \bibnamefont{and} \bibinfo{author}{\bibfnamefont{F.}~\bibnamefont{Wolf}},
  \bibinfo{journal}{Nucl. Phys.} \textbf{\bibinfo{volume}{B897}},
  \bibinfo{pages}{500} (\bibinfo{year}{2015}{\natexlab{a}}),
  \eprint{1503.07634}.

\bibitem[{\citenamefont{Blumenhagen
  et~al.}(2015{\natexlab{b}})\citenamefont{Blumenhagen, Damian, Font,
  Herschmann, and Sun}}]{Blumenhagen:2015xpa}
\bibinfo{author}{\bibfnamefont{R.}~\bibnamefont{Blumenhagen}},
  \bibinfo{author}{\bibfnamefont{C.}~\bibnamefont{Damian}},
  \bibinfo{author}{\bibfnamefont{A.}~\bibnamefont{Font}},
  \bibinfo{author}{\bibfnamefont{D.}~\bibnamefont{Herschmann}},
  \bibnamefont{and} \bibinfo{author}{\bibfnamefont{R.}~\bibnamefont{Sun}}
  (\bibinfo{year}{2015}{\natexlab{b}}), \eprint{1510.01522}.

\bibitem[{\citenamefont{Intriligator et~al.}(2006)\citenamefont{Intriligator,
  Seiberg, and Shih}}]{Intriligator:2006dd}
\bibinfo{author}{\bibfnamefont{K.~A.} \bibnamefont{Intriligator}},
  \bibinfo{author}{\bibfnamefont{N.}~\bibnamefont{Seiberg}}, \bibnamefont{and}
  \bibinfo{author}{\bibfnamefont{D.}~\bibnamefont{Shih}},
  \bibinfo{journal}{JHEP} \textbf{\bibinfo{volume}{04}}, \bibinfo{pages}{021}
  (\bibinfo{year}{2006}), \eprint{hep-th/0602239}.

\bibitem[{\citenamefont{Abe et~al.}(2007)\citenamefont{Abe, Higaki, Kobayashi,
  and Omura}}]{Abe:2006xp}
\bibinfo{author}{\bibfnamefont{H.}~\bibnamefont{Abe}},
  \bibinfo{author}{\bibfnamefont{T.}~\bibnamefont{Higaki}},
  \bibinfo{author}{\bibfnamefont{T.}~\bibnamefont{Kobayashi}},
  \bibnamefont{and} \bibinfo{author}{\bibfnamefont{Y.}~\bibnamefont{Omura}},
  \bibinfo{journal}{Phys. Rev.} \textbf{\bibinfo{volume}{D75}},
  \bibinfo{pages}{025019} (\bibinfo{year}{2007}), \eprint{hep-th/0611024}.

\bibitem[{\citenamefont{Kallosh and Linde}(2007)}]{Kallosh:2006dv}
\bibinfo{author}{\bibfnamefont{R.}~\bibnamefont{Kallosh}} \bibnamefont{and}
  \bibinfo{author}{\bibfnamefont{A.~D.} \bibnamefont{Linde}},
  \bibinfo{journal}{JHEP} \textbf{\bibinfo{volume}{02}}, \bibinfo{pages}{002}
  (\bibinfo{year}{2007}), \eprint{hep-th/0611183}.

\bibitem[{\citenamefont{Binetruy and Dudas}(1996)}]{Binetruy:1996uv}
\bibinfo{author}{\bibfnamefont{P.}~\bibnamefont{Binetruy}} \bibnamefont{and}
  \bibinfo{author}{\bibfnamefont{E.}~\bibnamefont{Dudas}},
  \bibinfo{journal}{Phys. Lett.} \textbf{\bibinfo{volume}{B389}},
  \bibinfo{pages}{503} (\bibinfo{year}{1996}), \eprint{hep-th/9607172}.

\bibitem[{\citenamefont{Blumenhagen et~al.}(2009)\citenamefont{Blumenhagen,
  Conlon, Krippendorf, Moster, and Quevedo}}]{Blumenhagen:2009gk}
\bibinfo{author}{\bibfnamefont{R.}~\bibnamefont{Blumenhagen}},
  \bibinfo{author}{\bibfnamefont{J.~P.} \bibnamefont{Conlon}},
  \bibinfo{author}{\bibfnamefont{S.}~\bibnamefont{Krippendorf}},
  \bibinfo{author}{\bibfnamefont{S.}~\bibnamefont{Moster}}, \bibnamefont{and}
  \bibinfo{author}{\bibfnamefont{F.}~\bibnamefont{Quevedo}},
  \bibinfo{journal}{JHEP} \textbf{\bibinfo{volume}{09}}, \bibinfo{pages}{007}
  (\bibinfo{year}{2009}), \eprint{0906.3297}.

\bibitem[{\citenamefont{de~Carlos
  et~al.}(2010{\natexlab{a}})\citenamefont{de~Carlos, Guarino, and
  Moreno}}]{deCarlos:2009fq}
\bibinfo{author}{\bibfnamefont{B.}~\bibnamefont{de~Carlos}},
  \bibinfo{author}{\bibfnamefont{A.}~\bibnamefont{Guarino}}, \bibnamefont{and}
  \bibinfo{author}{\bibfnamefont{J.~M.} \bibnamefont{Moreno}},
  \bibinfo{journal}{JHEP} \textbf{\bibinfo{volume}{01}}, \bibinfo{pages}{012}
  (\bibinfo{year}{2010}{\natexlab{a}}), \eprint{0907.5580}.

\bibitem[{\citenamefont{de~Carlos
  et~al.}(2010{\natexlab{b}})\citenamefont{de~Carlos, Guarino, and
  Moreno}}]{deCarlos:2009qm}
\bibinfo{author}{\bibfnamefont{B.}~\bibnamefont{de~Carlos}},
  \bibinfo{author}{\bibfnamefont{A.}~\bibnamefont{Guarino}}, \bibnamefont{and}
  \bibinfo{author}{\bibfnamefont{J.~M.} \bibnamefont{Moreno}},
  \bibinfo{journal}{JHEP} \textbf{\bibinfo{volume}{02}}, \bibinfo{pages}{076}
  (\bibinfo{year}{2010}{\natexlab{b}}), \eprint{0911.2876}.

\bibitem[{\citenamefont{Danielsson and Dibitetto}(2013)}]{Danielsson:2012by}
\bibinfo{author}{\bibfnamefont{U.}~\bibnamefont{Danielsson}} \bibnamefont{and}
  \bibinfo{author}{\bibfnamefont{G.}~\bibnamefont{Dibitetto}},
  \bibinfo{journal}{JHEP} \textbf{\bibinfo{volume}{03}}, \bibinfo{pages}{018}
  (\bibinfo{year}{2013}), \eprint{1212.4984}.

\bibitem[{\citenamefont{Blåbäck et~al.}(2013)\citenamefont{Blåbäck,
  Danielsson, and Dibitetto}}]{Blaback:2013ht}
\bibinfo{author}{\bibfnamefont{J.}~\bibnamefont{Blåbäck}},
  \bibinfo{author}{\bibfnamefont{U.}~\bibnamefont{Danielsson}},
  \bibnamefont{and}
  \bibinfo{author}{\bibfnamefont{G.}~\bibnamefont{Dibitetto}},
  \bibinfo{journal}{JHEP} \textbf{\bibinfo{volume}{08}}, \bibinfo{pages}{054}
  (\bibinfo{year}{2013}), \eprint{1301.7073}.

\bibitem[{\citenamefont{Danielsson and Dibitetto}(2014)}]{Danielsson:2013rza}
\bibinfo{author}{\bibfnamefont{U.}~\bibnamefont{Danielsson}} \bibnamefont{and}
  \bibinfo{author}{\bibfnamefont{G.}~\bibnamefont{Dibitetto}},
  \bibinfo{journal}{JHEP} \textbf{\bibinfo{volume}{05}}, \bibinfo{pages}{013}
  (\bibinfo{year}{2014}), \eprint{1312.5331}.

\bibitem[{\citenamefont{Dudas et~al.}(2007)\citenamefont{Dudas, Papineau, and
  Pokorski}}]{Dudas:2006gr}
\bibinfo{author}{\bibfnamefont{E.}~\bibnamefont{Dudas}},
  \bibinfo{author}{\bibfnamefont{C.}~\bibnamefont{Papineau}}, \bibnamefont{and}
  \bibinfo{author}{\bibfnamefont{S.}~\bibnamefont{Pokorski}},
  \bibinfo{journal}{JHEP} \textbf{\bibinfo{volume}{02}}, \bibinfo{pages}{028}
  (\bibinfo{year}{2007}), \eprint{hep-th/0610297}.

\end{thebibliography}

\end{document}